\documentclass[traditabstract]{aa}

\usepackage{graphicx}

\newcommand{\kms}{km\,s$^{-1}$}
\newcommand{\HI}{\ion{H}{i}}
\newcommand{\KI}{\ion{K}{i}}
\newcommand{\NaI}{\ion{Na}{i}}
\newcommand{\CaI}{\ion{Ca}{i}}
\newcommand{\CaII}{\ion{Ca}{ii}}
\newcommand{\FeI}{\ion{Fe}{i}}
\newcommand{\AlI}{\ion{Al}{i}}
\newcommand{\TiI}{\ion{Ti}{i}}
\newcommand{\TiII}{\ion{Ti}{ii}}
\newcommand{\CrI}{\ion{Cr}{i}}
\newcommand{\MnI}{\ion{Mn}{i}}
\newcommand{\SrII}{\ion{Sr}{ii}}

\begin{document}

\title{Gas phase atomic metals in the circumstellar envelope of IRC+10216
\thanks{Based on observations made with the VLT telescope and the UVES 
spectrograph at the European Southern Observatory in Paranal, Chile 
(ESO program 66.C-0220).
}}

\author{N.~Mauron\inst{1} \and P.J. Huggins\inst{2}}

\offprints{N.~Mauron}

\institute{GRAAL, CNRS and Universit\'e  Montpellier II,  
 Place Bataillon, 34095 Montpellier, France\\ 
 \email{mauron@graal.univ-montp2.fr}
\and 
Physics Department, New York University, 4 Washington Place, New York NY 10003, USA\\
\email{patrick.huggins@nyu.edu}
   }

\date{}

\abstract{
We report the results of a search for gas phase atomic metals in the
circumstellar envelope of the asymptotic giant branch carbon star
IRC+10216.  The search was made using high resolution ($\lambda
/\Delta \lambda = 50\,000$) optical absorption spectroscopy of a
background star that probes the envelope on a line of sight 35\arcsec\
from the center. The metal species that we detect in the envelope
include \NaI, \KI, \CaI, \CaII, \CrI, and \FeI, with upper limits for
\AlI, \MnI, \TiI, \TiII, and \SrII. The observations are used to
determine the metal abundances in the gas phase and the condensation
onto grains. The metal depletions in the envelope range from a factor
of 5 for Na to 300 for Ca, with some similarity to the depletion
pattern in interstellar clouds.  Our results directly constrain the
condensation efficiency of metals in a carbon-rich circumstellar
envelope and the mix of solid and gas phase metals returned by the
star to the interstellar medium.   The abundances of the uncondensed
metal atoms that we observe are typically larger than the abundances
of the metal-bearing molecules detected in the envelope.  The metal
atoms are therefore the major metal species in the gas phase and
likely play a key role in the metal chemistry.}

\titlerunning{Metals in the envelope of IRC+10216}
\authorrunning{N.~Mauron \& P.J.~Huggins}
\maketitle


\section{Introduction} 

Metals are important constituents of the interstellar medium. Abundant
metals such as calcium and iron are among the most heavily depleted
elements in dense clouds, and therefore form a significant component of
interstellar dust. The process of dust formation is not well
understood (e.g., Draine 2009), but much of the raw material is
supplied by mass loss from evolved stars, especially the asymptotic
giant branch (AGB) stars.

In the winds of oxygen-rich AGB stars, the dust is primarily in the
form of metal silicates, so that the metals play an important role in the
formation of dust and the mass loss process. In carbon-rich AGB stars,
the situation is less clear. The dust is believed to consist mainly of
graphite or amorphous carbon, and silicon carbide. The extent to which
metals contribute to dust formation, and the form in which they are
returned to the interstellar medium is essentially unknown.

In this paper, we report the first comprehensive search for gas phase
atomic metals in a carbon-rich circumstellar envelope. IRC+10216 (CW
Leo) is the nearest carbon star with a thick circumstellar envelope,
and serves as an archetype for the study of mass loss on the AGB.  The
star is relatively faint, $\sim$16.0~mag.\ in the $R$-band and much fainter at
shorter wavelengths because of obscuration by the envelope, but at
longer wavelengths the circumstellar dust and gas are seen in
emission, and are brighter than for any similar object. IRC+10216 has
therefore been intensively observed, with more than 50 molecular
species detected in the envelope, including several metal bearing
species (e.g., Olofsson 2005; Ziurys 2006a).

The technique that we use here to search for atomic metals in the
envelope of IRC+10216 is optical absorption spectroscopy, using a
background source of illumination. The star itself is too faint and
its spectrum too complex to serve as a useful source for detailed
study, although circumstellar C$_2$ and CN have been observed in this
way (Bakker et al.\ \cite{bakker97}).  There are, however, other stars
in the field. IRC+10216 is nearby, at a distance of $\sim 120$~pc
(e.g., Ramstedt et al.\ 2008), so the circumstellar envelope extends a
considerable angle on the sky. It is detected out to a distance of
$\sim 3\arcmin$ from the center in millimeter CO emission (Huggins et
al.\ 1988), and 9\arcmin\ in infrared dust emission observed with IRAS
(Young et al.\ 1993).  Although IRC+10216 is at a relatively high
galactic latitude ($l = 221\degr$, $b = +47\degr$), there are several
stars in this region of the sky that are candidates for background
sources, as seen in the wide field image in Fig.~1 of Mauron \&
Huggins (1999).

One of these stars, Star~6 in the UBV photometric sequence of the
field by Mauron et al.\ (2003), is well suited for absorption line
studies.  It is located behind the envelope at an angular offset of
35\arcsec\ from the center, and is bright enough for high resolution
spectroscopy. This star has been observed with the UVES spectrograph
at the VLT by Kendall et al.\ (2002). Their main objective was to
search for diffuse bands that might originate in the circumstellar
gas. No diffuse bands were found, but these authors noted deep
absorption lines of \NaI\ and \KI, which they attributed to
circumstellar gas.

Here we report a comprehensive search for metal lines in the
circumstellar envelope of IRC+10216 along this line of sight.  Our
objectives are to measure the degree of metal depletion
in this carbon rich-environment, and to determine the distribution
of solid and gas phase metals returned by the star to the interstellar
medium. Sect.~2 describes the observational material.  Sect.~3
presents the absorption lines, and Sect.~4 the derived column densities
and abundances. The results are discussed in terms of dust formation in
Sect.~5, and metal chemistry in Sect.~6. Our main conclusions are
given in Sect.~7.


\section{Observations}

The observations of the envelope of IRC+10216 were made using
absorption line spectroscopy with Star~6 (USNO\,0975-0633-6975) as the
background source of illumination.  Star~6 lies 35\arcsec\ from
IRC+10216 at position angle 165\degr. Its visual magnitude is
$V$=16.0, and it is the nearest star to the center of the envelope
suitable for high resolution spectroscopy.  The field is shown in
Fig.~1, where the envelope is seen in dust-scattered Galactic light.
From its magnitude and spectral type (type G), star~6 is at a
distance of $\sim 1400$~pc, well beyond IRC+10216, as discussed by
Kendall et al.\ (\cite{kendall02}).
 
The observations were made with the UVES spectrograph at the VLT, and
were previously used by Kendall et al.\ (2002) to search for diffuse
bands.  The data were obtained over seven nights in December 2000 and
January 2001, with an effective exposure of 4~hr at each wavelength.
The sky transparency was moderate to good, and the seeing was  0\farcs8.
The data consist of 4 spectra, each covering the complete range from
3\,000 to 10\,000~\AA\ with a resolving power of 50\,000 ($\Delta v =
6$~\kms). These spectra have been reduced with the ESO UVES pipeline
and have been summed after correction to the heliocentric reference
frame. All radial velocities given in this paper are heliocentric,
except where specified otherwise. The signal-to-noise ratio of the
final spectrum at 4200~\AA\ is $\sim 60$ per resolution element.

The heliocentric velocity of Star~6 measured from numerous
photospheric absorption lines is $+52.4\pm0.6$~\kms. For comparison,
the systemic heliocentric radial velocity of IRC+10216 is
$-$19.3~\kms, which is derived from millimeter observations of the
envelope ($v_{\rm LSR} = -26$~\kms, Loup et al.\ 1993). Thus a given
line in the envelope of IRC+10216 is well separated in velocity from
the same line formed in the photosphere of Star~6.  In addition, the
expansion velocity of the envelope is 14.1~\kms\ with a small
turbulent component (Huggins \& Healy 1986), so the absorption lines
in the envelope are expected to be wide ($\sim 30$~\kms). In
contrast, the weak stellar lines are relatively narrow, with widths
comparable to the instrumental profile, and the strong stellar lines have
characteristic damping profiles with broad wings.


\begin{figure}[!ht]
\resizebox{8cm}{!}{\rotatebox{-00}{\includegraphics{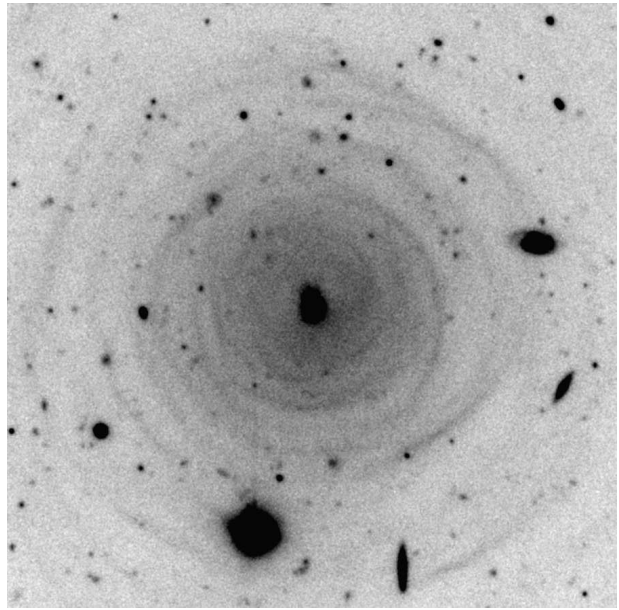}}}
\caption[]{$V$-band image of the circumstellar envelope of IRC+10216,
  made with the VLT. The field size is $90\arcsec
  \times90\arcsec$. The mass-losing carbon star is located at the
  center of the image, and Star~6 is the bright source near the
  bottom, 35\arcsec\ from the center. North is to the top, East to the
  left.}
\label{fig01}
\end{figure}


\begin{table}[!ht]
	\caption{Metal lines in the envelope of IRC+10216 }
        \label{table:1}
        \centering
	\begin{tabular}{lcccc}
	\noalign{\smallskip}
	\hline
	\hline
	\noalign{\smallskip}
Species & $\lambda_\circ$ & $f$-value & $W_{\lambda}^{\mathrm a}$  &   $N$    \\
        &      (\AA)      &     & (m\AA) & (cm$^{-2}$)      \\

	\noalign{\smallskip}
	\hline
	\noalign{\smallskip}
\NaI   & 5895.924  &  0.3180 &   625    & 4.6(14)$^{\mathrm b}$ \\
       & 5889.951  &  0.6311 &   625    &             \\
       & 3302.368  &  0.0090 &   230    &             \\
\noalign{\smallskip}
\AlI   & 3944.006  &  0.1134 &   $<$44  & $<$2.8(12)  \\
\noalign{\smallskip}
\KI    & 7698.974  &  0.3393 &   530    & 4.6(12)$^{\mathrm c}$ \\
       & 7664.911  &  0.6816 &   635    &             \\ 
       & 4044.143  &  0.0061 &   $<$29  &             \\	
\noalign{\smallskip}
\CaI   & 4226.728  &  1.7530 &   300    &  1.9(12)    \\
\noalign{\smallskip}
\CaII  & 3968.468  &  0.3145 &   255    &  7.0(12)    \\ 
       & 3933.663  &  0.6346 &   390    &             \\
\noalign{\smallskip}
\TiI   & 3635.462  &  0.2229 &   $<$49  & $<$1.9(12)  \\ 
\noalign{\smallskip}
\TiII  & 3383.768  &  0.3401 &   $<$69  & $<$2.0(12)  \\ 
\noalign{\smallskip}
\CrI   & 4289.716  & 0.0622  &     37:  & 1.4(12)$^{\mathrm d}$ \\ 
       & 4274.796  & 0.0839  &     49:  &             \\ 
       & 4254.332  & 0.1099  &     38:  &             \\ 
       & 3605.322  & 0.2248  &     27:  &             \\ 
       & 3593.482  & 0.2897  &     62:  &             \\
       & 3578.683  & 0.3663  &   $<$62  &             \\ 
\noalign{\smallskip}
\MnI   & 4034.483  & 0.0257  &  $<$33 & $<$3.1(12)    \\ 
       & 4033.062  & 0.0402  &  $<$33 &               \\ 
       & 4030.753  & 0.0565  &  $<$33 &               \\ 
\noalign{\smallskip}
\FeI   & 3859.911  &  0.0217 &  225   &  8.8(13)$^{\mathrm e}$  \\
       & 3824.444  &  0.0048 &   80   &               \\
       & 3719.934  &  0.0412 &  197   &               \\
       & 3440.606  &  0.0236 &  170   &               \\  
\noalign{\smallskip}
\SrII  & 4077.709  &  0.7010 & $<$28  & $<$2.7(11)    \\   
   
 \noalign{\smallskip}
 \hline
 \end{tabular}
\begin{list}{}{}
\item[$^{\mathrm{a}}$] Upper limits are 3$\sigma$.
\item[$^{\mathrm{b}}$] D-lines are saturated, $N$ based on the 3302~\AA\ line.
\item[$^{\mathrm{c}}$] Based on the 7699, 7665~\AA\ doublet.
\item[$^{\mathrm{d}}$] : indicates 2--3$\sigma$ features; $N$
  based on weighted mean.
\item[$^{\mathrm{e}}$] Based on the 3860, 3720~\AA\ lines which have
  better S/N. 
\end{list}
\end{table}


Table~1 lists the metal lines that we searched for in the envelope of
IRC+10216. Column~2 gives the laboratory wavelength in air
($\lambda_\circ$) for each transition, and column~3 gives the
oscillator strength ($f$) from Morton (1991, 2000).  The line list includes
the strongest ground-state transitions for the species that are
potentially observable in the wavelength range covered.  These lines
are seen in interstellar clouds and/or in circumstellar environments
with similar physical conditions.

At long wavelengths, the spectrum of the background source, Star~6, is
relatively free of photospheric lines, so an absorption line arising
in the intervening circumstellar envelope is straightforward to
identify.  At shorter wavelengths, the spectrum is more crowded with
photospheric lines, and absorption by the circumstellar envelope is
often more difficult to identify. In these cases we use a template
technique to extract the envelope signal. We first make a least
squares fit to the spectrum of Star~6 using a template spectrum covering
a region of $\sim 15$~\AA\ around (but excluding) the expected
envelope line. We then use this as the effective continuum to
search for residual absorption from the envelope.

One form of template that we tried was based on the library of
synthetic spectra from Coelho et al.\ (2005), but these did not produce a
good match to the photospheric spectrum.  A second approach that was
successful, was using a scaled solar spectrum. By chance the spectrum
of Star~6 is very similar to that of the Sun, and a template based on
scaling the solar spectrum gives a close match to the stellar
spectrum.  The solar spectrum that we use here was published by
Delbouille et al.~(1973) and covers the wavelength region
3\,000--10\,000~\AA\ with a step size of about 0.0125 \AA. In order to
fit the stellar spectrum $F_*(\lambda)$, we shift the solar spectrum
$F_{\odot}(\lambda)$ to match the
radial velocity, re-bin it to match the resolution, and scale the flux
according to:
\[ F_*(\lambda) = \alpha F_{\odot}(\lambda) + \beta\,, \] 
where $\alpha$ and $\beta$ are constants that are determined from the
least-squares fit to the local region of spectrum under consideration.

The use of this template as the effective continuum reveals spectral
features of the circumstellar envelope that are not otherwise easily
observed.  Examples of the application of this technique are given in
Sects. 3.3 and 3.4.


\begin{figure}   [!t]
\resizebox{8.2cm}{!}{\includegraphics{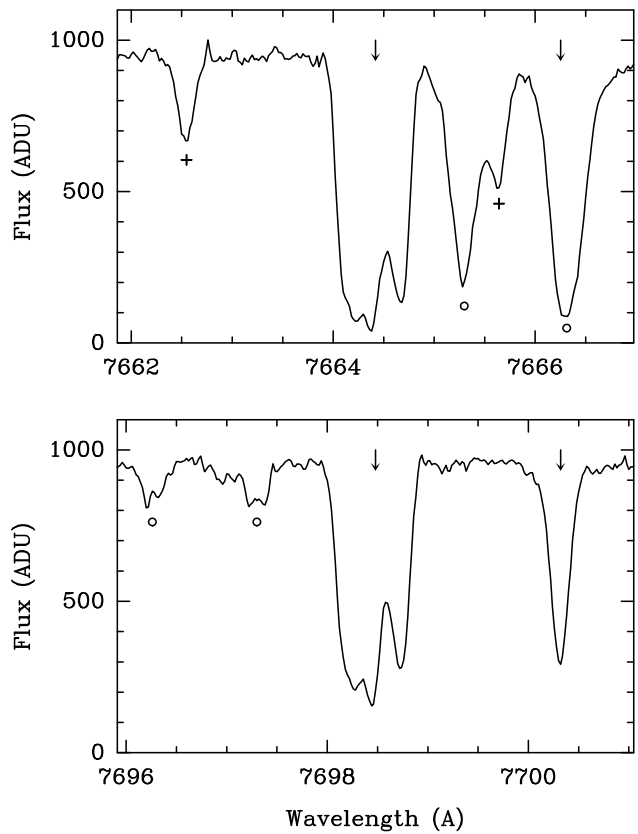}}
\caption[]{Spectra of the \KI\ doublet at 7664.91~\AA\ (\emph{upper
panel}) and 7698.97~\AA\ (\emph{lower panel}). In each panel, the
arrow on the right marks the photospheric line ($V_{\mathrm r} =
+52.4$~\kms), and the arrow in the center marks the envelope
absorption of IRC+10216 ($V_{\mathrm r} = -19.3$~\kms). Other
photospheric lines are marked $+$, and telluric lines are marked
$\circ$.
 }
\label{fig02}
\end{figure}


 \section{Envelope absorption lines}

 \subsection{\KI}

Fig.~2 shows the observed spectral regions covering the \KI\
doublet. The 7665~\AA\ line is in the upper panel and the 7699~\AA\
line is in the lower panel. Strong photospheric \KI\ lines of Star~6
are present at the stellar radial velocity of $V_{\mathrm r} =
+52.4$~\kms, which is marked with an arrow at the right of each panel.
Also present are two weaker photospheric lines (marked $+$) and four
telluric lines (marked $\circ$).

The dominant feature in the middle of each spectral region is strong
\KI\ absorption from the envelope of IRC+10216. This can be
unambiguously ascribed to the circumstellar envelope. It is centered
near the systemic radial velocity of IRC+10216 at $-19.3$~\kms\
(marked with an arrow in the figure) and has the broad width expected
from the expanding envelope.  With the low interstellar absorption in
this region of the sky ($E_{B-V} \la 0.03$), any interstellar \KI\
would contribute $\la$ a few percent of the strong lines observed,
based on the survey of interstellar \KI\ by Chaffee \& White (1982).

We also searched for the much weaker ground state \KI\ line at
4044~\AA\ which has an $f$-value $\sim110$ times less than the
7664~\AA\ line. The 4044~\AA\ line was not detected, and the upper
limit is given in Table~1.

Although the \KI\ doublet lines formed in the envelope are broad, they
are clearly composed of several components.  The presence of the
components is important because it affects the saturation of the lines
(see Sect.~4). Assuming gaussian broadening, we found that a best fit
synthesis taking into account the instrumental profile requires four
components to match the line shapes. A preliminary fit with
unconstrained parameters gave consistent estimates for the radial
velocities of the individual components in each line. We then fixed
the velocities at their mean values, and fit the profiles by varying
the relative column densities of the components and their line
broadening b-values, which we constrained to be the same for both
lines. The final mean parameters are given in Table 2,  where the
strength is the column density of each component, relative to the
strongest component.  The uncertainty in the velocities is $\la
0.5$~\kms\ and the relative strengths found from the two lines are the
same to within $\la 5$\%.  Thus the results from the two lines are in
good agreement, as expected from their similar line profiles. 

 To illustrate the quality of the multi-component fit, Fig.~3 shows the
results of a synthesis of the 7699~\AA\ line using the parameters of
Table~2. It can be seen that the model provides an excellent fit to the
observational data.

We interpret the multiple component profiles of the \KI\ lines as the
effects of the multiple shell structure in the envelope (Mauron \&
Huggins 1999, 2000; see also Fig.~1), where the line of sight passes
through regions of enhanced density. Although the gaussian line shapes
give a good fit to the spectra, we do not know the detailed velocity
distribution of the shells along the line of sight, so we cannot
reliably determine the gas density contrast between the shell and
inter-shell regions. However, the stronger components are clearly
separate, and suggest that the contrast is a least a factor of a
few. This is consistent with a large shell inter-shell contrast in the
dust density derived from images in dust-scattered light by Mauron \&
Huggins (2000).


\begin{figure}   [!t]
\resizebox{8.2cm}{!}{\includegraphics{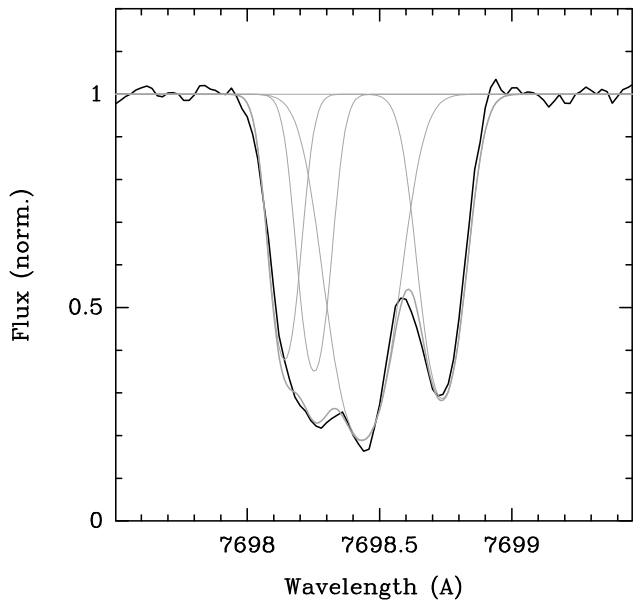}}
\caption[]{Comparison of the observed (black) and synthesized (grey)
  profiles of the \KI\ 7698.97~\AA\ line, using the
  parameters given in Table~2. The individual components are also
  shown.  }
\label{fig03}
\end{figure}



\begin{table}[bht]

	\caption{Parameters of the \KI\ components in IRC+10216} 
        \label{table:2}
        \centering
	\begin{tabular}{cccc}
	\noalign{\smallskip}
	\hline
	\hline
	\noalign{\smallskip}
Component    &  Strength    &  $b$   &     $V_{\mathrm r}$   \\
             &   (rel.)     & (\kms) &      (\kms)   \\
	\noalign{\smallskip}
	\hline
	\noalign{\smallskip}

 1   &   0.27  &   2.8   &  $-$32.4  \\
 2   &   0.32  &   2.9   &  $-$28.0  \\
 3   &   1.00  &   5.5   &  $-$21.0  \\
 4   &   0.51  &   3.9   &  $-$09.2  \\

 \noalign{\smallskip}
 \hline
 \end{tabular}
\end{table}



\begin{figure}   [!t]
\resizebox{8.2cm}{!}{\includegraphics{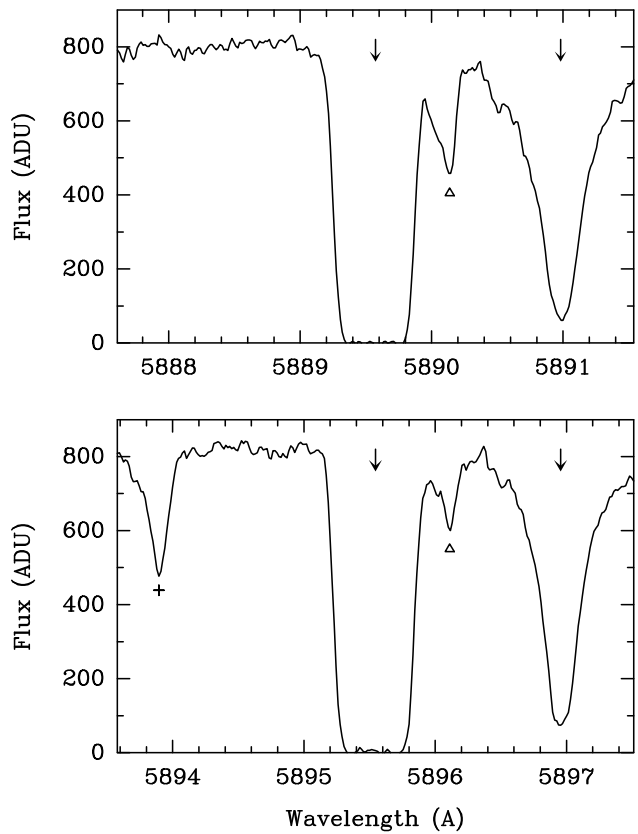}}
\caption[]{ Spectra of the \NaI\ doublet at 5889.95~\AA\ (\emph{upper
panel}) and 5895.92~\AA\ (\emph{lower panel}). Arrows mark the
photospheric and envelope components, as in Fig.~2. The triangles mark
an interstellar \NaI\ component (at $V_{\mathrm r} \sim +9.5$~\kms).  
}
\label{fig04}
\end{figure}


 \subsection{\NaI}

Fig.~4 shows the spectral regions covering the \NaI\ D lines. The
5890~\AA\ line is in the upper panel and the 5896~\AA\ line is in the
lower panel.  The photospheric \NaI\ lines of Star~6 (marked with
arrows at the right in each panel) are strong, and there is another
weaker photospheric line in the lower panel.  The dominant feature in
the middle of each spectrum is \NaI\ absorption from the envelope of
IRC+10216, centered near the systemic velocity. It can be seen that
the circumstellar lines are extremely strong.  All the components are
highly saturated and the residual intensities are close to zero across
both lines.

There is an additional, weak \NaI\ component that appears in both
spectra near $+$9.5~\kms\ (marked with triangles) which we identify as an
interstellar component along this line of sight. Similar weak \NaI\
absorption near this velocity is reported by Kendall et al.\ (2002)
along lines of sight to two other stars in this region of the sky, at
angular distances of 153\arcsec\ and 2.5\degr\ from IRC+10216.

We also searched for the much weaker ground state \NaI\ line at
3302~\AA, which has an $f$-value $\sim 70$ times less than the
5890~\AA\ D line.  The 3302~\AA\ line lies in a crowded, low
signal-to-noise part of the spectrum, but the template approach
(Sect.~2) reveals a $\sim$5 sigma detection, and the equivalent width
is given in Table~1. The signal-to-noise ratio is too low to show any
details of the profile.


\begin{figure}   [!t]
\resizebox{8.2cm}{!}{\includegraphics{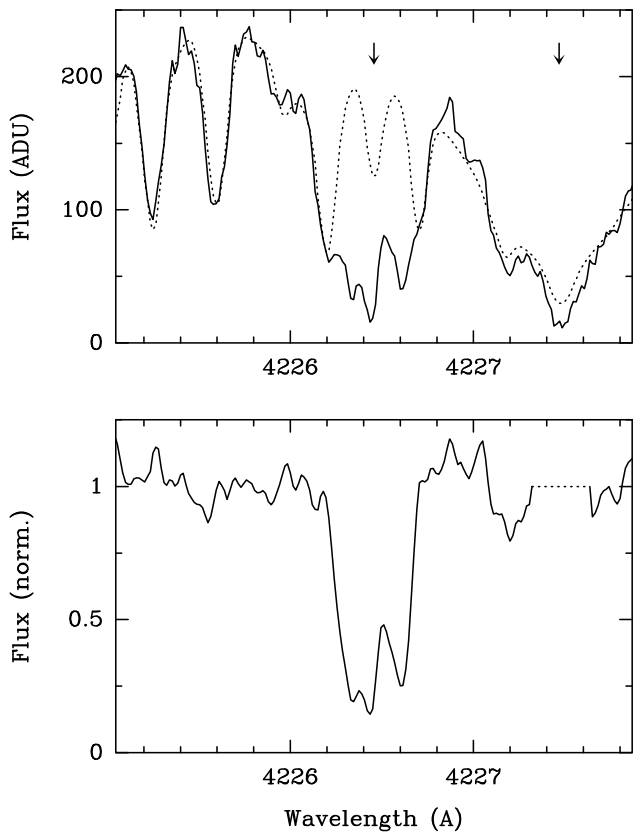}}
\caption[]{ Spectrum of the \CaI\ line at 4226.73~\AA.  \emph{Upper
 panel:} The observed spectrum (full line) and fitted template (dotted
 line).  Arrows indicate the photospheric and envelope components as in
 Fig.~2. \emph{Lower panel:} Spectrum normalized to the template. The
 dotted line replaces the spectrum over the photosperic core region.
}
\label{fig05}
\end{figure}



\begin{figure}   [!t]
\resizebox{8.2cm}{!}{\includegraphics{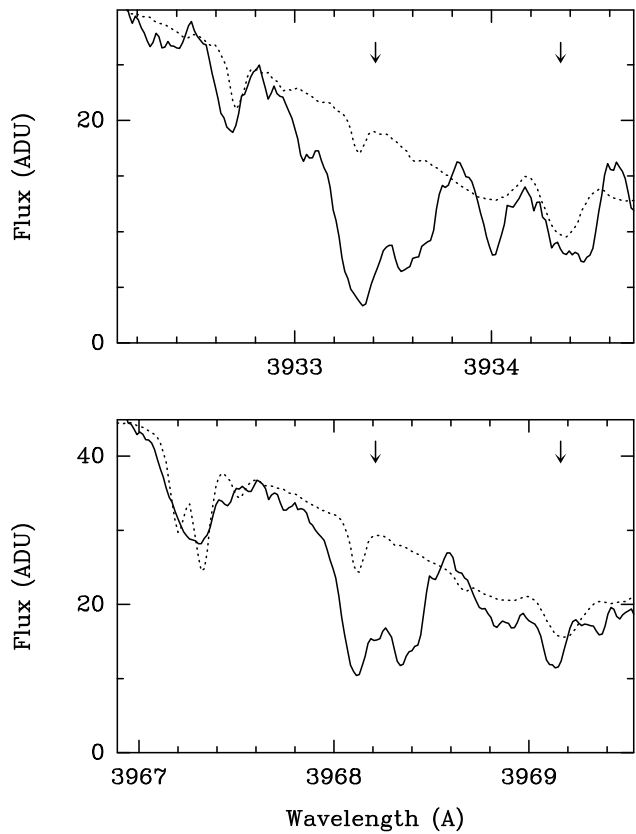}}
\caption[]{Spectra of the \CaII\ H and K doublet at 3933.66~\AA\ (\emph{upper
panel}) and 3968.46~\AA\ (\emph{lower panel}). Arrows mark the
photospheric and envelope components, as in Fig.~2. The solid lines
show the observed spectra, the dotted lines show the template spectra.
}
\label{fig06}
\end{figure}


 \subsection{\CaI\ and \CaII}
 
Fig.~5 shows the spectral region around the \CaI\ line at 4226~\AA.
The line falls in a very crowded region of the spectrum and provides a
good illustration of the template method.

The solid line in the upper panel shows the observed spectrum, and the
envelope absorption is not immediately apparent.  The dotted line in
the upper panel shows the stellar template, scaled to match the
observed spectrum.  It can be seen that it gives a good overall fit to
the data, and reveals the excess absorption from the envelope. The
lower panel of Fig.~5 shows the envelope line, using the fitted
template as the effective continuum. The small wavelength range around
the core of the photospheric \CaI\ line has been masked (with the
horizontal dotted line) because the effective continuum is poorly
defined at the low intensity of the line core. The main deviation from
a flat continuum in the normalized spectrum is caused by slight
differences between the \CaI\ line profiles of Star~6 and the Sun,
which probably arise from slightly different surface gravities.

It can be seen that the template fitting technique recovers the \CaI\
absorption in the envelope very effectively. The line shows a profile
with narrow components similar to those seen in the \KI\ lines.

We also detected \CaII\ absorption in the H and K lines shown in
Fig.~6. The 3933~\AA\ line is in the upper panel, and the 3968~\AA\
line is in the lower panel. For these spectral regions the crowding by
photospheric lines is not severe, and the envelope absorption can be
seen in the direct spectra. The adjacent \CaII\ photospheric lines are
very strong, and their blue wings form the local (tilted) continuum
for the circumstellar absorption. In the photospheric line core, the
template is not a good fit on account of differences in the core
reversals between Star 6 and the Sun.

The \CaII\ line profiles show approximately the same profiles as the
\KI\ lines but are somewhat broader, by $\sim$ 5--10~\kms (FWHM),
suggesting additional absorption.  This could be an interstellar
contribution, or additional circumstellar absorption. Compared to the
neutral lines, the ionized \CaII\ lines sample the line of sight
through the envelope to larger distances from the central star, and
the kinematics of these outer regions have never previously been
observed.


\begin{figure}   [!t]
\resizebox{8.2cm}{!}{\includegraphics{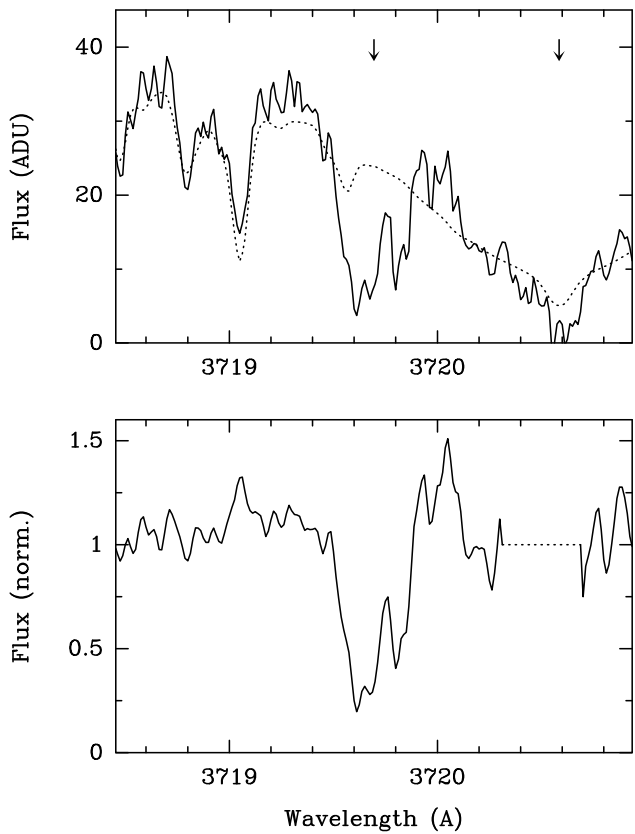}}
\caption[]{ Spectrum of the \FeI\ line at 3719.93~\AA. Details as in Fig.~5. 
}
\label{fig07}
\end{figure}



\begin{figure}   [!t]
\resizebox{8.2cm}{!}{\includegraphics{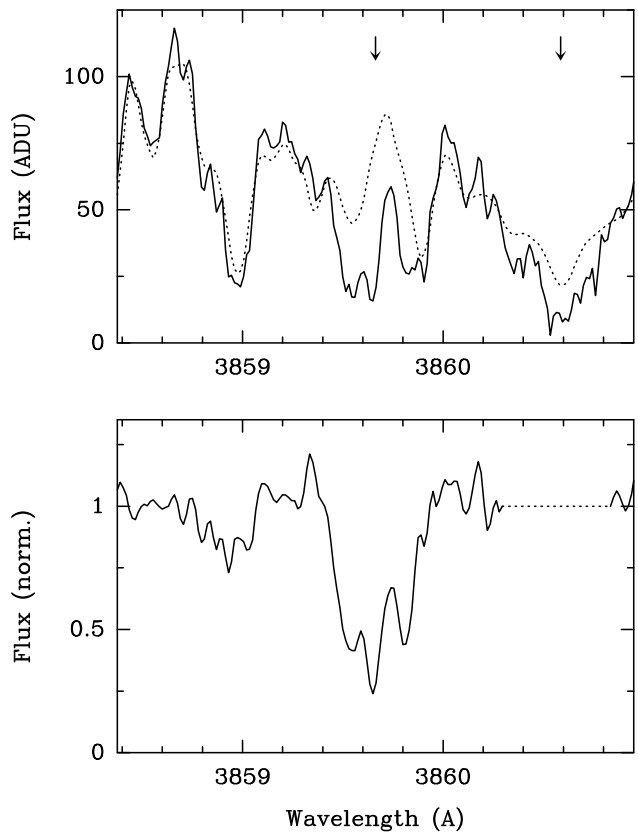}}
\caption[]{ Spectrum of the \FeI\ line at 3859.91~\AA. Details as in Fig.~5. 
}
\label{fig08}
\end{figure}


\subsection{\FeI}

Figs.~7 and 8 show the spectral regions around the \FeI\ lines at
3720~\AA\ and 3860~\AA, respectively.  These fall in relatively
crowded regions of the photospheric spectrum, and the template is
needed to determine the envelope absorption. The upper panel in each
figure shows the observed spectrum (solid line) and the fitted stellar
template (dotted line).  The lower panel shows the normalized spectrum
with the envelope absorption centered near the systemic velocity of
IRC+10216.  The signal-to-noise ratio of these spectra is lower than
those discussed above, and the template fit is affected by differences
in the solar and stellar spectra.   The limited quality of the fit is
probably responsible for the fact that the equivalent width of the
3720~\AA\ line is slightly less than that of the 3860~\AA\ line, even
though the f-value is larger (see Table~1).  Nevertheless the lines
are well detected, and the multi-component character of the envelope
absorption is similar to that seen in the lines of the other species.

We also searched for two additional \FeI\ lines at 3824~\AA\ and
3441~\AA.  The 3824~\AA\ line is weaker on account of a significantly
lower $f$-value, and is marginally detected at $\sim$4 sigma. The
3441~\AA\ line lies at shorter wavelengths where the spectrum is
poorer. It is detected with a strength comparable to the \FeI\ lines
shown in Figs.~7 and 8, but with a lower signal-to-noise ratio.

\subsection{Other lines}

Searches were also made for lines of several other species, including
\AlI, \MnI, \CrI, \TiI\ and \TiII, and \SrII, as listed in
Table~1. With the exception of \CrI, no significant envelope
absorption was detected.  For \CrI, weak ($\sim$2--3$\sigma$) features
are seen at the wavelengths of most of the accessible lines. When
appropriately weighted by the oscillator strengths and noise levels
they provide an overall 4-sigma detection of this species.

\section{Abundances in the envelope}

\subsection{Column densities }

The equivalent widths ($W_\lambda$) of the absorption lines observed
in the envelope are listed in column~4 of Table~1. For the
non-detected lines, we list 3-sigma limits given by:
\[ W_\lambda  < 3 \, \sigma \Delta \lambda / \sqrt n\, , \] where $\sigma$
is the noise level and $n$ is the number of resolution elements in the
line width $\Delta \lambda$.

The column density ($N$) derived for each species is given in
column~5 of Table~1.  It is based on the equivalent widths and the
$f$-values given in the table.  For the weak lines, the column density
is directly related to the equivalent width by the optically thin
formula:
\[  N = 1.13\times10^{20}\, W_\lambda / f \lambda^2, \] 
where $ W_{\lambda}$ and $\lambda$ are in \AA\ units.

For the stronger lines, where the optical depths are larger, we
determine the relation between $W_{\lambda}$ and $N$ using the
multi-component model used to fit the high signal-to-noise profiles of
the \KI\ lines (Table~2).  With this model we find that the \KI\ lines
are moderately saturated; the column densities derived from the 7699
and 7665~\AA\ lines are larger than the optically thin values by
factors of 1.7 and 2.3, respectively. The column densities from the
two lines agree to within 20\%, and are consistent with a limit of
$N({\KI}) \la 3\times10^{13}$~cm$^{-2}$ determined from the upper
limit to the optically thin 4044~\AA\ \KI\ line.   For \NaI, the D-lines
are extremely optically thick, and therefore insensitive to the column
density, although the very low residual intensity across the lines
yields a lower limit of $N(\NaI) \ga
$1--$2\times10^{13}$~cm$^{-2}$.   Fortunately, we also detect the
optically thinner \NaI\ line at 3302~\AA\ line, and we use this with
the multi-component saturation curve to determine the column density
of \NaI\ given in Table~1. The result is consistent with the lower
limit from the D-lines.

The profiles of the ionized lines might be expected to differ from
those of the neutral lines because the absorption can occur at
different locations along the line of sight. This does not affect the
column density estimates for \SrII\ and \TiII\ because the lines are
very weak (below the detection level) and are therefore optically
thin. For the stronger \CaII\ lines, the profiles are found to be
similar to the \KI\ profiles, but as noted in Sect.~3.3, they are
slightly broader with an additional contribution to the absorption. We
therefore model the \CaII\ lines as the sum of two contributions; the
fit of the \KI\ multi-component model to the main part of the profile,
and an additional, optically thin component to give the total
equivalent width. The additional component contributes 32\% and 17\%
of the equivalent widths (22\% and 13\% of the derived column
densities) for the 3933 and 3968~\AA\ lines, respectively. We assume
that the additional absorption is circumstellar, but since it is
relatively small, the results do not depend sensitively on this
assumption.

\subsection{Column density of hydrogen}

Although there are no direct observations of hydrogen along the line
of sight, we can use estimates of the mass loss rate of the envelope
to determine the hydrogen column density. Since the early work by Kwan
\& Hill (1977) there have been numerous estimates of the mass loss
rate based on millimeter CO observations, and the results are fairly
consistent when the different distances, CO abundances, dust-gas
heating rates, and He content are taken into account.  For a distance
of 120~pc, we adopt a mass loss rate $\dot{M}_{\mathrm H}$ (in
hydrogen) of $1.25 \times 10^{-5}$~$M_{\odot}$~yr$^{-1}$
 (corresponding to a total mass loss rate of $1.75 \times
10^{-5}$~$M_{\odot}$~yr$^{-1}$), based on the analysis of Sch\"{o}ier
\& Olofsson (2001) approximately corrected for the effects of He.
This value is consistent with other recent estimates.

For $\dot{M}_{\mathrm H} =1.25 \times 10^{-5}$~$M_{\odot}$~yr$^{-1}$,
the column density of hydrogen $N$(H), where $N$(H) = $N$(\HI) +
$2N$(H$_2$), is $1.3 \times 10^{21}$~cm$^{-2}$ along the line of sight
35\arcsec\ from the center.  The uncertainty is a factor $\sim 2$,
which results from uncertainties in $\dot{M}$. In addition to a smooth
decrease of the column density with distance from the center, there
are other variations caused by the multiple shell structure in the
envelope.  From an analysis of scattered light images (e.g., Fig.~1) we
find typical variations in the dust column density of $\pm20\%$, and
somewhat smaller variations near the region of Star 6.   These
variations are much smaller than the density contrast of the shells
because the column density averages the density along the line of sight
through several shell and inter shell regions.  The envelope structure
is therefore not a major source of uncertainty for $N$(H).

\subsection{Ionization fraction}

The line of sight passes through regions of the circumstellar envelope
where the metals are partially ionized. Since most of the observed
lines arise from single stages of ionization, we need to consider
ionization corrections in order to determine the total gas phase
abundances. We estimate the corrections from the relative column
densities of the ionization stages of each metal, obtained using the
photo-ionization model of an expanding envelope discussed by Glassgold
\& Huggins (1986).

In the model, neutral atoms (which may result from the dissociation of
molecules) emerge from the dense, shielded, inner envelope and are
photo-ionized by the ambient interstellar radiation field. We
calculate the ionization fraction as a function of radius, and use
this to calculate the ionization fraction along the line of sight. The
ionization of each element is governed by equation 4.9 of Glassgold \&
Huggins (1986), with no contribution from chromospheric radiation. The
ionization depends on the photo-ionization rate (given by the
interstellar rate and the envelope shielding) and recombination. In
solving the ionization equations we use the interstellar
photo-ionization rates and recombination rates from P\'equignot \&
Aldrovandi (1986), except for Cr and Sr which are not included in
their compilation; for these the photo-ionization rates are from
Glassgold \& Huggins (1986) and the recombination rates from Bernat
(1976). The radial dependence of the electron abundance is adopted
from Cordiner et al.\ (2007), although recombination is relatively
unimportant except for Al. For the shielding of interstellar radiation
in the envelope we use the standard carbon dust model and dust-to-gas
ratio of Cherchneff et al.\ (1993) and the shielding function of
Morris \& Jura (1983).

There are a number of uncertainties in the parameters of the model,
the most important being the strength of the ambient radiation field
(which determines the interstellar photo-ionization rates) and the
shielding in the envelope (which includes uncertainties in the dust
parameters, the mass loss rate, and the envelope
geometry). Fortunately we observe both the neutral and ionized column
densities of Ca, and we use this to set the ionization level in the
model. For the nominal parameters given above, the predicted ratio
$N(\CaII)/N(\CaI)$ is a factor 3.6 larger than observed. This is fair
agreement considering that the ionization in the envelope has not
previously been constrained in this way. However, we can do much
better, by fine tuning the ionization level in the envelope to fit the
Ca ionization exactly. This can be done by adjusting the ambient
radiation field (by a factor of 0.46) or the shielding optical depth
(by a factor 1.5).  Both variations are within their respective
uncertainties.  For specificity, we adopt the reduced radiation field
to calculate the ionization of the other elements. The resulting
ionization corrections ($C_{\mathrm i}$ = $N$({\sc i} + {\sc
ii})/$N$({\sc i}) for neutral species, and $N$({\sc i} + {\sc
ii})/$N$({\sc ii}) for singly ionized species), are given in column~4
of Table~3.  Adjusting the shielding instead, by the amount given
above, produces essentially the same ionization corrections.

The ionization correction for \AlI\ in Table~3 is much larger than for
the other metals because of its relatively high photo-ionization
rate. Our observations are therefore not very sensitive probes of the
column density of Al in the gas phase because the Al is nearly
completely ionized along the line of sight.  The ionization
corrections for the other neutral species are much smaller. For
example, Na is predominantly neutral.  This is in contrast to typical
interstellar clouds with a similar column density of hydrogen. In the
circumstellar envelope, the characteristic outflow time at 35\arcsec\
is shorter than the photo-ionization time given by Glassgold \&
Huggins (1986). Hence, even without dust shielding, the Na atoms are
expected to be largely neutral, as given in the table.

\subsection{Abundances}

Except for Ca and Ti, the total gas phase column densities of the
metals are determined from the observed column densities and the
ionization corrections. The results are given in column 5 of Table~3.
Although the $N(\CaII)/N(\CaI)$ ratio was used to determine the
ionization level in the envelope, we use the sum of the ionization
stages to obtain the total Ca column density, independent of the
ionization. Similarly, the limit for Ti is based on the observed
limits for \TiI\ and \TiII, and so is independent of the ionization.

The gas phase abundances of the metals relative to hydrogen ($X$) are
determined from the total column densities and the value of
$N({\mathrm H})$ from Sect.~4.2, and are given in column~6 of Table~3.
For reference, the solar values ($X_\odot$) are given in column~7 of
the table, taken from Lodders (\cite{lodders03}).  Comparison of the
envelope and solar abundances shows that there are large deficiencies
in the gas phase abundances in the envelope, which vary from metal to
metal.

\begin{table*}[!ht]

	\caption{Abundances in the envelope of IRC+10216} 
        \label{table:3}
        \centering
	\begin{tabular}{lccccccccc}
	\noalign{\smallskip}
	\hline
	\hline
	\noalign{\smallskip}
El. & Ion & $N$     &  $C_{\mathrm i}^{\mathrm{a}}$  &   $N$({\sc i}+{\sc ii})   &    $N$({\sc
	i}+{\sc ii})/$N$(H) & $X_\odot$   & $\log_{10} \delta$  &
	$\log_{10} \delta_{7027}$ & $\log_{10} \delta_{\zeta\,\mathrm{Oph}}$  \\
 &  &  (cm$^{-2}$) &   & (cm$^{-2}$) \\
	\noalign{\smallskip}
	\hline
	\noalign{\smallskip}

Na  & \sc{i}   & $\phantom{<}$4.6(14) &  1.22  & 5.6(14) & 4.2($-$7) &
2.00($-$6) & $-$0.68 & $-$0.06 & $-$0.95 \\

\noalign{\smallskip}
Al  & \sc{i} & $<$2.8(12)           & 3.5(3)& $<$9.8(15) &   $<$7.3($-$6)
& 2.88($-$6) & $\ldots^{\mathrm{c}}$  & $\ldots$ & $\ldots$ \\

\noalign{\smallskip}
K   & \sc{i} & $\phantom{<}$4.6(12) &  1.60  & 7.4(12) &  5.5($-$9) &
1.29($-$7) & $-$1.37 &  $-$0.17 &  $-$1.09\\ 

\noalign{\smallskip}
Ca  & \sc{i} & $\phantom{<}$1.9(12) &  4.70  & 8.9(12)$^{\mathrm{b}}$
&  6.6($-$9) & 2.19($-$6) & $-$2.52 & $-$0.75 & $-$3.73  \\

    & \sc{ii} & $\phantom{<}$7.0(12) &  1.27  & $\ldots$ & $\ldots$
&    $\ldots$ & $\ldots$ & $\ldots$ & $\ldots$\\

\noalign{\smallskip}
Ti  & \sc{i} & $<$1.9(12)           &  4.26  & $<$3.9(12)$^{\mathrm{b}}$   &  $<$2.9($-$9) &
8.32($-$8) & $<$\,$-$1.45 & $\ldots$  & $-$3.02\\

      & \sc{ii}  & $<$2.0(12)       &  1.31  & $\ldots$    &  $\ldots$
& $\ldots$  & $\ldots$ & $\ldots$ & $\ldots$  \\

\noalign{\smallskip}
Cr  & \sc{i} & $\phantom{<}$1.4(12) &  10.6 & 1.5(13) & 1.1($-$8) &
4.47($-$7) & $-$1.60 &   $\ldots$ &  $-$2.28 \\

\noalign{\smallskip}
Mn  & \sc{i} & $<$3.1(12)           &  2.21  &  $<$6.9(12) &
$<$5.1($-$9) & 3.16($-$7) &  $<$\,$-$1.79 &  & $-$1.45 \\

\noalign{\smallskip}
Fe  & \sc{i} & $\phantom{<}$8.8(13) &  2.58  & 2.3(14) & 1.7($-$7) &
2.95($-$5)  & $-$2.24 &  $-$1.68 & $-$2.27 \\

\noalign{\smallskip}
Sr & \sc{ii} & $<$2.7(11)           &  3.82  & $<$1.0(12) &  $<$7.6($-$10) &
8.13($-$10)& $\ldots^{\mathrm{c}}$  &   $\ldots$ &   $\ldots$ \\

 \noalign{\smallskip}
 \hline
 \end{tabular}
\begin{list}{}{}
\item[$^{\mathrm{a}}$] Ionization correction $C_{\mathrm i} = N({\sc I} + {\sc
  II}) /N({\sc I})$ for neutral species and  $N({\sc I} + {\sc II})
  /N({\sc II})$ for singly ionized species. 
\item[$^{\mathrm{b}}$] Based on observed values of $N$({\sc I}) + $N$({\sc
  II}), independent of $C_{\mathrm i}$.
\item[$^{\mathrm{c}}$] Upper limit on $X > X_\odot$, no useful limit
  on $\delta$. 
\end{list}

\end{table*}


\section{Dust condensation}

Several processes affect the state of the circumstellar material as it
moves from the stellar photosphere out through the envelope into the
interstellar medium.  In order of increasing distance from the star
these processes include: dust condensation in conditions of
approximate thermodynamic equilibrium; gas phase chemical reactions;
photo-dissociation of the molecules; and eventual photo-ionization of
the atomic constituents in the outer envelope (e.g., Gilman 1969;
Tsuji 1973; McCabe et al.\ 1979; Huggins \& Glassgold 1982; Lafont et
al.\ 1982). In the carbon-rich environment of IRC+10216, the main
component of the dust is amorphous carbon with a minor component of
SiC (e.g., Martin \& Rogers 1987). The state of other elements,
especially the refractory metals, is not well understood (e.g., Turner
1995). Hence our observations of the gas phase metals in the outer
envelope provide new constraints on the gas phase chemistry and the
dust condensation.

\subsection{Observed depletions}

Comparison of the envelope abundances with the solar abundances in
Table~3 shows that most of the gas phase metal atoms in the
envelope are ``missing''. It is most unlikely that they are in the
form of gas phase molecules. The fractional abundance of even the most
extreme case of Ca, where $X(\mathrm{Ca})/X_{\odot} \sim 3\times
10^{-3}$, is an order of magnitude \emph{larger} than the largest
fractional abundance of any metal bearing molecule detected outside of
the core region (see Table~4). In addition, most molecules are
dissociated closer to the star than the 35\arcsec\ offset of our line
of sight.  It is therefore reasonable to infer that the atomic
abundances observed are good approximations to the total gas phase
metal abundances in the envelope, and that the missing atoms are
depleted onto dust grains.

In column (8) of Table~3 we give the conventional measure of depletion
$\log_{10} \delta$, where $\delta = X/X_{\odot}$ is the depletion
factor. The observational limits for the abundances of Al and Sr are
$\ga$ the solar values, so in these cases we have no significant
limits for $\delta$, although Sr is an $s$-process element and may be
enhanced in IRC+10216 \emph{and} depleted. 

Based on the measured depletions, there are two immediate
conclusions. First, the metals in this carbon-rich archetype are
primarily in the form of solids, and this is the dominant form
returned to the interstellar medium. Second, in spite of the depletion, a
significant residue of metallic atoms remains in the gas phase and varies
from metal to metal.

\subsection{Condensation and adsorption}

There are no detailed predictions for the depletion of metals in
IRC+10216, but there are some important considerations that bear on
the issue. A commonly used approach to the condensation of solids in
circumstellar envelopes is the assumption of thermodynamic equilibrium
in the warm, dense, inner envelope, where the chemical time scales are
rapid compared with the expansion time scale. Under these conditions
the formation of solid particles is controlled by the condensation
temperature of the primary condensate of each species.

For a carbon-rich envelope, the condensation sequence depends somewhat
on the C/O ratio and the gas pressure. The following sequence, for C/O
= 1.1 and a pressure of $10^{-6}$~bar (from Lodders \& Fegley 1995,
updated for Fe by Lodders \& Fegley 1999) is representative: C
(1670~K), TiC (1640~K), SiC (1460~K), FeSi (1230~K), AlN (1170~K), CaS
(1150~K), MgS (960~K), with other more volatile metals such as Na and
K at lower temperatures. Cr and Mn probably form sulphides but their
location in the sequence is uncertain.  Thus the metals Ti, Fe, Al,
Ca, and then K and Na, are expected to be removed from the gas phase
successively. Those with lower condensation temperatures are less
likely to go to completion because of the decreasing density with
radius.

This qualitative picture is largely consistent with the observed
depletion pattern. The observations provide a firm upper limit on the
gas phase abundance of Ti (from the \TiI\ and \TiII\ lines);
Fe and Ca are strongly depleted, although Ca is more depleted than Fe (in
reverse order to the condensation sequence); and Na and K are less
depleted.

The simplifying assumption of thermodynamic equilibrium is not a
complete physical theory of condensation because kinetic effects must
play a role near freeze-out.  In addition, once formed, the grains can
act as sites for the adsorption of gas phase species further out in
the envelope (Jura \& Morris 1985). The adsorption depends on the
binding energy of the species to the grain surface, and on the
sticking probability ($p$), which is essentially unknown. Using the
analysis of Jura \& Morris (1985), Turner (1995) finds that for
IRC+10216, a volatile species such as K is depleted by adsorption to a
fractional abundance of 0.84 (for $p=0.1$) and 0.17 (for $p=1$), and a
more refractory species such as Al is depleted to a fractional
abundance of 0.05 (for $p=0.1$) and $10^{-10}$ (for $p=1$). Thus
adsorption of metals may be as important as the initial condensation.

Our observations of a residual atomic component in the gas phase
constrain the efficiency of both condensation and adsorption. Below
about 1150~K, the phase diagrams of Lodders \& Fegley (1995) show that
Ca is essentially completely removed from the gas phase, but our
observations show that even for this refractory metal there is a
residual component in the gas phase. Metal bearing molecules with Na
and K (albeit with low fractional abundances) are seen in the core
region (see Sect.~6) but not in the extended envelope. This may be the
result of adsorption. On the other hand, adsorption of the more
refractory species to levels of $10^{-10}$ are clearly ruled out by
the observations.

We conclude that current theoretical ideas are qualitatively in accord
with our observations of metal depletion, but improved models with
specific quantitative predictions are needed to discriminate the
underlying processes.

\subsection{Comparison with PNe}

Carbon-rich planetary nebulae (PNe) are the immediate descendants of
carbon-rich AGB stars, in which the circumstellar material has
undergone major changes. The gas has been photo-ionized, and the dust
grains are exposed to intense radiation fields and high ($\sim
10^4$~K) temperatures. A comparison of depletions in AGB stars and PNe
may therefore reveal some aspects of grain evolution.

Element abundances have been extensively measured in the ionized gas
in PNe. The abundances are, however, subject to systematic, and
sometimes, large uncertainties, and relatively few metals have
accessible lines covering the appropriate stages of ionization.  For
comparison with IRC+10216 we focus on NGC~7027, which is one of the
most intensively studied, carbon-rich PNe. NGC~7027 is relatively
young and still surrounded by a substantial envelope of molecular gas
(Cox et al.\ \cite{cox02}). The circumstellar conditions before the
formation of the nebula were therefore somewhat similar to the current
state of IRC+10216.

There have been numerous abundance analyses of NGC~7027. We have taken
the abundances of Na, K, Ca, and Fe from recent, comprehensive studies
by Keyes et al.\ (1990), Middlemass et al.\ (1990), Bernard Salas et
al.\ (2001), and Zhang et al.\ (2005), and we give the corresponding
depletions in column (9) of Table~3. For species in common the
depletions have been averaged. Even with these state-of-the-art
analyses, the differences in abundances between the different studies
range up to a factor of $\sim 3$.

Comparison of the depletions in IRC+10216 and NGC~7027 reveals some
significant differences. First, the metals Na and K are much less
depleted in the ionized nebula, where they are close to the solar
values. Second, Fe and Ca are still significantly depleted in the
nebula but, evidently less than in the circumstellar envelope. These
results suggest the following evolutionary effects in the transition
from AGB star to PN: the nearly complete evaporation of the volatile
species Na and K, and the partial erosion of more refractory species
Ca, and possibly Fe.  Further evidence for this view comes from the
study of NGC~7027 by Kingdon et al.\ (1995), who argue that Ca is
depleted by more than 2 orders of magnitude near the periphery of the
nebula (as in IRC+10216), but is much less depleted near the center of
the ionized nebula.

The abundances in NGC~7027 are fairly similar to other carbon-rich
PNe, e.g., the Fe depletion is similar to the typical Fe depletion
found in a sample of low ionization PNe by Delgado Inglada et al.\
(2009).  The trends noted here may therefore be a general
characteristic of the evolution of dust from the AGB to PNe.

\subsection{Comparison with the ISM}

It is also of interest to compare our results for IRC+10216 with
depletions in the ISM.  We take the line of sight towards $\zeta$~Oph
as representative of the ISM, bearing in mind that the overall level
of IS depletions varies with location but the general pattern
remains the same.  The depletions towards $\zeta$~Oph, from Savage \&
Sembach (1996), are listed in column (10) of Table~3.

In spite of the different physical and chemical environments that lead
to grain formation in IRC+10216 and the ISM, it can be seen that the
depletion patterns are qualitatively similar. Na and K are the least
depleted, at comparable levels; Mn, Fe, and Cr are probably similar
although the detailed pattern may differ; and Ca is the most depleted
in both data sets.

In studies of the ISM it is found that element depletions correlate
with the (oxygen-rich) condensation temperature of the element-bearing
solid. This was interpreted in terms of grain formation at high
temperatures in the winds of mass-losing giants (Field 1974). More
recent studies indicate that a significant component of the dust in
the ISM is formed in situ.  The correlation with the condensation
temperature may therefore reflect some aspect which is shared by the
formation process in the ISM.  The similarity of the pattern that we
find in IRC+10216 may have some bearing on this question, and deserves
further attention.

\begin{table}[!ht]

\caption{Metal bearing molecules in IRC+10216} 
\label{table:4}
\centering
\begin{tabular}{lcccc}
\noalign{\smallskip}
\hline
\hline
\noalign{\smallskip}
Species   & $\theta^{\mathrm a}_{\mathrm r} $  & $X^{\mathrm b}_{\mathrm{mol}}$  & $X_{\mathrm{mol}}/X_\odot$  & Ref.  \\
          &  (\arcsec)  \\
\noalign{\smallskip}
\hline
\noalign{\smallskip}

NaCl   &   2.5    &  1.3($-$09)  &  6.3($-$4)  &  1 \\

NaCN   &   2.5    &  6.8($-$09)  &  3.4($-$3)  &  1 \\

\noalign{\smallskip}

KCl    &    2.5   &  3.8($-$10)  &  2.9($-$3)  &  2 \\ 
 
\noalign{\smallskip}

AlF   &    2.5    &  4.6($-$08)  &  1.6($-2$) &  1  \\

AlCl  &    2.5    &  2.8($-$08)  &  9.8($-$3) &  1  \\

AlNC  &   5--15   &  2.3($-$10)  &  7.9($-$5) &  3  \\

\noalign{\smallskip}

MgNC  &   10--20  &  6.3($-$09)  & 1.8($-$4)  &  1  \\

MgCN  &   10--20  &  2.9($-$10)  & 8.0 ($-$6) &  4  \\

 \noalign{\smallskip}
 \hline
 \end{tabular}
\begin{list}{}{}
\item
$^{\mathrm a}$ Source radius.
\item
$^{\mathrm b}$ Abundance relative to H, for $d=120$~pc,  $\dot{M}(\mathrm{H})=1.25 \times 10^{-5}$~$M_{\odot}$~yr$^{-1}$.
\item
References: (1) Highberger \& Ziurys\ (\cite{highberger03}). (2)
Assuming $X({\rm{KCl}}) = 0.3 X({\rm{NaCl}})$ Cernicharo \&
Gu\'elin (\cite{cernicharo87}). (3) Ziurys et al.\
(\cite{ziurys02}). (4) Assuming $X({\rm{MgCN}}) = 0.045 X({\rm{MgNC}})$ Ziurys et al.\ (\cite{ziurys95}).   
\end{list}
\end{table}


\section{Metal chemistry}

The circumstellar envelope of IRC+10216 exhibits a remarkable gas
phase chemistry, with more than 50 molecular species detected to date,
mainly through their rotational transitions at millimeter wavelengths.
The majority of the molecules are formed of the abundant elements H,
C, N, and O, several include Si and S, but a small number unexpectedly
include metals, Na, K, Mg, and Al. The first of these were discovered
by Cernicharo \& Gu\'elin (1987).  Our observations of gas phase metal
atoms in IRC+10216 constrain some aspects of the metal chemistry in
the envelope.

The metal bearing molecules detected in the envelope are listed in
Table~4. In addition to these, numerous other metal bearing species
have been searched for at comparable levels, but not detected (e.g.,
Turner 1995). There are also other metal species whose rotational
frequencies have not yet been measured in the laboratory (e.g., Ziurys
2006b).

For comparison with the atomic abundances reported here, we list
updated abundances for the metal molecules in column (3) of
Table~4. These are based on the column densities given in the
references in the table, but are derived for the distance and mass
loss rate used in this paper, and are relative to hydrogen. The
fractional abundance $X_{\mathrm{mol}}/X_{\odot}$, is the fraction of
each metal in a particular molecular species. $\theta_{\mathrm r}$ is
the measured, or inferred, angular radius of the molecular
distribution, adopted from the references.

The metal bearing molecules divide naturally into two groups. The
first is confined to the dense core region represented by
$\theta_{\mathrm r} \la 2.5\arcsec$ in the table.  The second, which
includes only Mg and Al bearing molecules, is found in distinct shells
in the envelope, in the photo-dissociation region. The general level
of incorporation of the metals into molecules is very small in both
groups.

In the outer envelope, the fractional abundances of the metal bearing
molecules are $\la 0.02$~per cent, while the atoms that we detect have
fractional abundances of 0.3--20~per cent. This we used in Sect.~5.1
to justify the assumption that the observed atoms are the dominant gas
phase metal carriers along the line of sight.

Even in the core region the molecules do not seem to be dominant. The
observed fraction of Na in the core in the form of  molecules is $\sim
0.4$~per cent, and we detect 20~percent in the form of atoms farther
out. Similarly, the fraction of K in the core in the form of molecules
is $\sim 0.3$~per cent, and we detect 4~per cent in atoms farther
out. For these species the numbers are consistent with the idea that
apart from condensation or adsorption onto dust grains, the dominant
gas phase species is atomic throughout the envelope.  There is lack of
specific molecular information for the other metals, but the absence
of detected molecular species argues that this applies to them as
well.

Our finding that residual metals are present in the envelope in the form of
neutral atoms and ions provides an observational foundation for
understanding certain aspects of the metal chemistry.  For example, it
has been proposed that metal cyanides are formed by reactions of metal
ions with cyanopolyyenes (Dunbar \& Petrie 2002), and the model of
Cordiner \& Millar (2009) shows that this can account for the observed
abundance of MgCN if enough Mg$^+$ is present in the gas phase. We
have not observed Mg, but we expect that it behaves like the other
metals. Its condensation temperature is significantly less than that
of Ca and Fe, so it is likely to be less depleted in the envelope. The
gas phase abundances of the metals that we observe are given in column
(6) of Table~3.  These may undergo reactions with cyanopolyyenes and
other abundant neutral molecules, such as unsaturated hydrocarbons,
and lead to a variety of metal bearing species.  The largest
abundances that we observe in the gas are those of Na and Fe (whose
large depletion is balanced by its high cosmic abundance). Thus Fe
offers interesting possibilities for a potentially observable Fe
chemistry.

\section{Conclusions}

The observations of IRC+10216 reported in this paper represent the
first comprehensive study of atomic metals in a carbon-rich
circumstellar envelope. We detect lines of Na, K, Ca, Cr, and Fe, and
obtain upper limits for Al, Ti, Mn, and Sr.  Combined with a simple
model of the ionization, the observations provide estimates of the gas
phase metal abundances in the outer envelope.

The results show that the metals, especially Ca and Fe, are
significantly depleted onto dust grains in the circumstellar envelope,
and this is the dominant form returned to the ISM. The depletion
pattern has some similarity with depletion in the ISM, and is roughly
consistent with expectations of dust condensation in a carbon-rich
envelope.

Although the metals are depleted in the envelope, atomic metals in the
form of neutral atoms and ions appear to be the major metal species in
the gas phase. As such, they likely play a key role in the metal
chemistry of the envelope.

\begin{acknowledgements}
We thank Dr.~K.~Lodders for helpful information on dust condensation. We also
thank an anonymous referee for helpful comments.
This work is supported in part by the National Science Foundation, 
grant AST 08-06910 (PJH).
\end{acknowledgements}




  \end{document}